\documentstyle[aps,preprint]{revtex}

\begin{document}

\title{ Decisions, Decisions:  Noise and its Effects on Integral Monte Carlo
Algorithms}

\author{J. D. Doll}
\address{Department of Chemistry,
Brown University,
Providence, RI 02912}

\author{D. L. Freeman}
\address{Department of Chemistry,
University of Rhode Island,
Kingston, RI 02881}

\maketitle
\begin{abstract}

	In the present paper we examine the effects of
noise on Monte Carlo algorithms, a problem
raised previously by Kennedy and Kuti (Phys. Rev. Lett.
{\bf 54}, 2473 (1985)).  We show that the
effects of introducing unbiased noise into the
acceptance/rejection phase of the conventional
Metropolis approach are surprisingly modest, and,
to a significant degree, largely controllable.  We
present model condensed phase numerical applications
to support these conclusions.
\end{abstract}
\pacs{PACS numbers: }
\section {Introduction}  As emphasized by Ulam in his
autobiography\cite{one}, the problem of decision
making in a statistical environment is an interesting
issue.  When procedures or input are corrupted
by ``noise,'' the resulting outcomes will in general
differ from their zero noise counterparts.  Such
modifications may range from inconsequential to extreme
depending on the details of the algorithm
and noise involved.

	Questions related to the spirit of this general
issue arise when we consider the effects of noise
on integral Monte Carlo techniques\cite{two}.  In these
methods ``decisions''
are made at various stages
according to a relatively simple set of rules.  Our
concern in the present paper is how the inclusion
of various types of noise into these decision making
steps alters the predictions of such algorithms.
In view of its practical importance, we focus attention
in the present discussion on the venerable
``Metropolis'' procedure\cite{three}.

	Monte Carlo algorithms in general are among the
most widely utilized numerical procedures in
science\cite{four}.  They have become, for example, a
primary tool for the analysis of many-body
phenomena.  The techniques are broadly applicable to
both classical\cite{five}
 and quantum-mechanical\cite{six} problems.

	Beyond their immediate numerical utility, these
statistical methods have served to promote the
development of new approaches to a diverse range of
important problems.  In the general field of
optimization, for instance, the simulated annealing
method\cite{seven} has emerged as a valuable technique
for function minimization.  This approach attempts to
find the minimum of a specified function by
recasting the problem as one of finding the zero ``temperature''
structure of a fictitious classical
system.  The ``potential energy'' of this system is taken
to be the function whose minimum is being
sought.  As emphasized by Press, et al.\cite{eight}, the
resulting method is a relatively robust one in that the
thermal fluctuations involved assist in avoiding local minima.
Related ``quantum annealing''
approaches have also been reported recently\cite{nine,ten,eleven}.

	The typical use of of the integral Monte Carlo
methods considered here is to obtain statistical
estimates of generalized averages that arise in various
physical applications.  Integral Monte Carlo
methods, in general, approximate such averages with statistical
estimates based on a finite set of
quadrature points drawn randomly from the probability
distribution functions in question.  The
Metropolis method\cite{three}, in particular, generates
these points by devising a random walk that visits
each configuration space location with a probability that
is proportional to its associated
probability.  If ``unbiased,'' these estimates can be made
arbitrarily exact by increasing the scale of
the random walk.  Methods suitable for treating ``rare event''
situations\cite{twelve} and for coping with
``sparse'' distributions\cite{thirteen,fourteen} are
described elsewhere.

	In typical integral Monte Carlo applications, the
distribution function involved is known with
certainty.  In other applications, however, the information
necessary to implement the Monte Carlo
procedure is available only numerically and only with limited
precision\cite{fifteen}.  It is thus important to
clarify the effects of such ``noise'' on the basic algorithms
involved if we are to utilize conventional
methods in these instances.

 	The outline of the remainder of present paper is as
follows:  Starting with a brief review of the
Metropolis method, we present in Section II a number of
results concerning the influence of noise
on this procedure.  In Section III we document the principal
conclusions of Section II with
prototypical numerical applications related to the study of
the dynamics of condensed-phase
quantum-mechanical systems.

\section{Formal Developments}  Detailed balance is at the
core of the Metropolis algorithm.  In
particular, the requirement that
\begin{equation}
\rho(A)P(A\rightarrow B) = \rho(B)P(B\rightarrow A)
\label{eq:1}
\end{equation}
is a sufficient condition to assure that a random walk
based
on the transitional rules, $P(X\rightarrow Y)$,visits
points, Y, in
proportion to their associated equilibrium densities,
$\rho(Y)$.  Rewritten slightly, the detailed balance
constraint
requires that the transition rules underlying the Monte Carlo
random walk satisfy
\begin{equation}
{P(A\rightarrow B)\over P(B\rightarrow A)} = {\rho(B) \over \rho(A)}.
\label{eq:2}
\end{equation}
There are many possible choices for transition probabilities
consistent with Eq. (2).  The original
Metropolis method\cite{three} takes
\begin{equation}
P(A\rightarrow B) = M({\rho_B / \rho_A}),
\label{eq:3}
\end{equation}
where the $M(x)$ is the ``Metropolis'' function,
\begin{equation}
M(x) = min(1,x).
\label{eq:4}
\end{equation}
For later purposes, it is convenient to take $M(x)$ to be
zero if $x$
is negative.  It is easy to verify that
\begin{equation}
M(x)/M(1/x) = x,
\label{eq:5}
\end{equation}
in accordance with the detailed balance.

	For typical problems, well established procedures exist
to generate the random walks specified
by Eq. (3).  The elemental building block for such walks is the
combination of randomly generated
``trial'' moves followed by ``acceptance/rejection'' decisions.
Descriptions of these procedures are
given elsewhere\cite{two,three}.  The essential points for the
present discussion are: (1) the decisions involved
are the mechanism by which detailed balance is enforced,
and (2) the decisions are based on
equilibrium density ratios.

       When we no longer know the required density ratios
precisely but
rather
possess only noisy, statistical estimates of the
required information, it
is not immediately obvious how best to proceed.  One
obvious approach is to
ignore the presence of noise and simply implement the
Metropolis procedure
in its original form.  If we do so, however, we have
no assurance the
resulting method will satisfy detailed balance.  In
particular, the
transition probability for such an approach,
$P_0(A\rightarrow B)$, can be shown to be,
\begin{equation}
P_0(A\rightarrow B) = \int dr M(r)f(r).
\label{eq:6}
\end{equation}
where  $f(r)$ denotes the probability that the B/A
density ratio for the step
assumes the value $r$.  Equation (6) obviously
reduces to Eq. (3) for
probability distributions sharply peaked about
the true ratio,
$\rho(B)/\rho(A)$,
In general, however, $P_0$ differs from its zero
noise counterpart and the
guarantee of detailed balance is lost.  The
influence of noise on the
Metropolis method can thus be explored by
examining its effects on the
transition probability (Eq. (6)) and on the
associated detailed balance
ratios (Eq. (2)).  Because of its importance
in practical applications,
we assume in what follows that $f(r)$ in Eq. (6)
is a normalized gaussian
with a mean equal to the exact density ratio.

        To explore the questions raised above,
we show in Figs. (1a) and
(1b) plots of the transition probabilities
produced by Eq. (6).  These were
obtained assuming gaussian noise of fixed absolute
(Fig. (1a)) and relative
widths (Fig. (1b)) of a particular magnitude (0.2)
in the density ratio.
For a given absolute noise, we see in Fig. (1a)
that $P$ and $P_0$ are
qualitatively similar, differing only in the regions
where the density
ratio is near zero or unity.  The widths of the
regions where differences
are observed are approximately the width of $f(r)$.
  It is easy to see from
the convolution structure of Eq. (6) that this is
a general result.
Explicitly, the average involved in Eq. (6) modifies
the resulting
transition probability relative to the zero
noise limit only in regions
where the Metropolis function is neither constant
nor purely linear over
the width of the gaussian noise.  For this reason,
the effects of a fixed
relative error in the acceptance/rejection density
ratio (Fig. (1b)) are
confined to the density ratio region near unity.

        While it is informative to examine the
behavior of the individual
transition probabilities themselves (Fig. (1)),
it is important to
emphasize that the primary measure of the effect
of noise on the
Metropolis
method is its consequence for detailed balance.
In Figs. (2a) and (2b) we
display plots of $P_0(x)/P_0(1/x)$ for a range of
density ratios, $x$, for the
systems examined in Fig. (1).  Breakdowns
in detailed balance are signaled
in such a plot by deviations from linearity.
In contrast to the behavior
of the individual transition probabilities in
Fig. (1a), we see in Fig.
(2a) that effects of absolute noise in the
density region near unity tend
to cancel from the detailed balance condition.
Evaluating Eq. (6)
analytically for noise of a given width, $\sigma$,
it is straightforward to
show that for small widths the effects of noise
are minimal over the density
ratio range $\sigma\le x\le 1/\sigma$.
{}From Fig. (2b) we see that relative
noise in the acceptance/rejection ratio has
little effect on the detailed
balance ratio in any region.  As will be documented
below, these
conclusions remain valid for a surprisingly large
range of absolute and
relative noise.

        The simple analysis presented above suggests
that the Metropolis
method is
robust with respect to the inclusion of unbiased
noise to its decision
making steps.  Even rather sizeable relative
errors in the
acceptance/rejection ratio appear to have little
effect on detailed
balance and hence on the overall performance of
the method.  While absolute
errors
have somewhat more pronounced consequences, their
effects are restricted
to an identifiable class of trial moves.
Specifically, we have seen above
that fixed noise enters the Metropolis method
principally thorough
decisions involving attempted moves between
points of grossly dissimilar
equilibrium probability.  Such moves can be
largely avoided in practice
by utilizing a trial displacement length scale
that is small relative to
that characterizing the interaction potential.
In the zero noise limit, the
selection of such length scales is purely a
question of efficiency and
any sensible choice will suffice.  We see,
however, that with the addition
of a specified absolute noise that smaller
length scales are preferable.
Their use will assure that the preponderance
of the acceptance/rejection
decisions will be made in situations where
the effects of noise are
essentially inconsequential.

\section{Numerical Example}  We present below
a brief numerical example chosen to illustrate
the potential use of the ideas discussed in
the previous section in study of quantum dynamics.  We
examine a model system\cite{sixteen} that
consists of a single moving particle trapped
on a line between two
space-fixed particles (separated by a distance L)
that act as a Rcage.S  The particle whose dynamics
is of interest is assumed to interact with the
fixed atoms through pairwise Lennard-Jones
potentials.  These potentials as well as the cage
dimensions can be varied to probe the behavior of
various prototypical systems.

	We focus in the following on a quantity
that plays a central role in describing the spectroscopy
and transport in finite temperature dynamical systems,
the ``self correlation'' function\cite{seventeen}.
For single
particle systems, this quantity can be defined as
\begin{equation}
G(y,t) = \langle\delta(y - (x'-x))\rangle,
\label{eq:7}
\end{equation}
where the brackets in Eq. (7) denote an average
over the probability
distribution $\rho(x)P(x't/x)$.
Here $\rho(x)$ describes the equilibrium density
as a function of
moving atom position while $P(x't/x) $
denotes to the conditional probability that a system
initially at position $x$ at time zero is at position
$x'$ a time $t$ later.  Generalizations to many-body
situations are straightforward.  The self correlation
function probes the likelihood of dynamical
displacements of a specified size in a finite temperature
system.  Quantum-mechanical expressions for the
required equilibrium density and dynamical
transition probability are available\cite{eighteen}
in terms of various complex temperature density matrix
elements.

	Our approach in what follows will be to
evaluate Eq. (7) for representative systems via Monte
Carlo sampling of the initial and final positions.
For molecular systems, the required equilibrium
tasks can be performed exactly in general.  We thus
sample initial positions in what follows in the
zero noise limit.  The information required to perform t
he dynamical sampling task, however, is
typically not available directly, but must itself be
obtained through a separate calculation.  To
simulate its possible presence in such calculations,
we add noise to the final position sampling
process as discussed below.

	Figure (3) displays $G(y,t)$ results for the
present model where potential and cage parameters
were selected to mimic the neon system
$(\epsilon/k_B = 35.6 K, \sigma = 2.749 \AA,
L = 2(2^{1/6}\sigma))$.  The
necessary quantum-mechanical matrix elements
were obtained using NMM
methods\cite{nineteen}.  The
Trotter index in the NMM calculations was taken
to be 128, more than adequate to assure
numerical convergence.  The particular results
shown in Fig. (3) correspond to a temperature of 51
K and correspond to  the zero noise limit.

	Certain general features of $G(y,t)$ are
evident in Fig. (3).  Classically, the self correlation
function approaches a delta function limit as
$t\rightarrow 0$.  We see in Fig. (3), however, that quantum-
mechanical effects produce a broadening of this
zero time classical behavior.  As time increases
near $t = 0$, particles move from their starting
locations increasing in the width of the corresponding
displacement distribution.  Ultimately, the
``caging effects'' of the fixed atoms in the model limit the
range of motion and produce the oscillatory structure
visible in Fig. (3).  Were the caging potential
harmonic, the structure would contain a single frequency
both classically and quantum-
mechanically.  Since the potential involved is
appreciably anharmonic, however, the temporal
structure obtained in Fig. (3) contains several
frequency components.  Quantum-mechanically,
these arise from the interplay between the system's
vibrational energy levels while classically they
reflect the energy dependence of the vibrational period.

	Figures (4) and (5) illustrate the effects of
including unbiased gaussian noise of a specified
relative and absolute magnitude, respectively, to the
sampling of the dynamical transition
probability for the system studied in Fig. (3).  As
expected, we see in Fig. (4) that the effects of
including relative noise are minimal for situations
where the signal to noise ratio is greater than 2:1.
Also as expected, we see in Fig. (5) that the effects
of including absolute noise are somewhat more
pronounced.  When utilized in the Metropolis procedure
with step sizes chosen by the usual
criterion (50\% acceptance), such noise produces
noticeable systematic errors.
These effects are largely eliminated, however, when
the displacement
length scale for the trial moves is reduced as
discussed in Section II.

\section{Summary}

	We have considered the effects of adding
noise to the decision making steps of the Metropolis
procedure.  Our investigation suggests that this
procedure is a relatively robust one with respect to
the addition of unbiased noise in the acceptance/
rejection probability ratio.  Specifically, noise of a
specified relative size has little effect on
detailed balance and hence on the functioning of the
method.  While the consequences of a fixed
absolute noise are potentially more severe, we find that
such effects are restricted to a particular
class of trial moves and are hence relatively easy to
minimize in practice.

\section{Acknowledgments}

	The authors would acknowledge support
through the National Science Foundation grant NSF-
CHE-9203498.  DLF also wishes to acknowledge
the donors of the Petroleum Research Fund of
the American Chemical Society for partial
support of this work.

\newpage
\section*{Figure Captions}

\begin{enumerate}

\item
Figure (1a) compares the usual Metropolis function (line
with discontinuous slope)
with the transition probability obtained from Eq. (6)
as a function of the trial
probability ratio.  Results shown were obtained
using gaussian noise with a fixed
standard deviation of 0.2.  Figure (1b) is the same
as (1a), except that the standard
deviation for the noise involved was of a specified
relative fraction (20\%) of the
probability ratio.

\item
Forward and reverse detailed balance quotients for the
systems described in Fig. (1)
as a function of the trial probability ratio.
Breakdowns in linearity signal deviations
from detailed balance.

\item
The self correlation function (Eq. (7))
for the Lennard-Jones cage problem
discussed in Section III.  System parameters
were those appropriate for neon.

\item
The self correlation function for the neon cage
problem in Fig. (3) at a specified
time (80,000 a.u.).  Results shown were obtained
from Eq. (7) with various
amounts of relative noise (0\% - 80\%) added to the trial probability ratio.

\item
The self correlation function for the neon
cage problem in Fig. (3) at a specified
time (80,000 a.u.).  Zero noise results (exact)
are compared with two calculations
involving a fixed level of noise (0.4).  In the
``noisy'' calculations, trial step sizes in
the final position sampling in Eq. (7) were
adjusted to produce 50\% and 80\%
acceptance, respectively.
\end{enumerate}
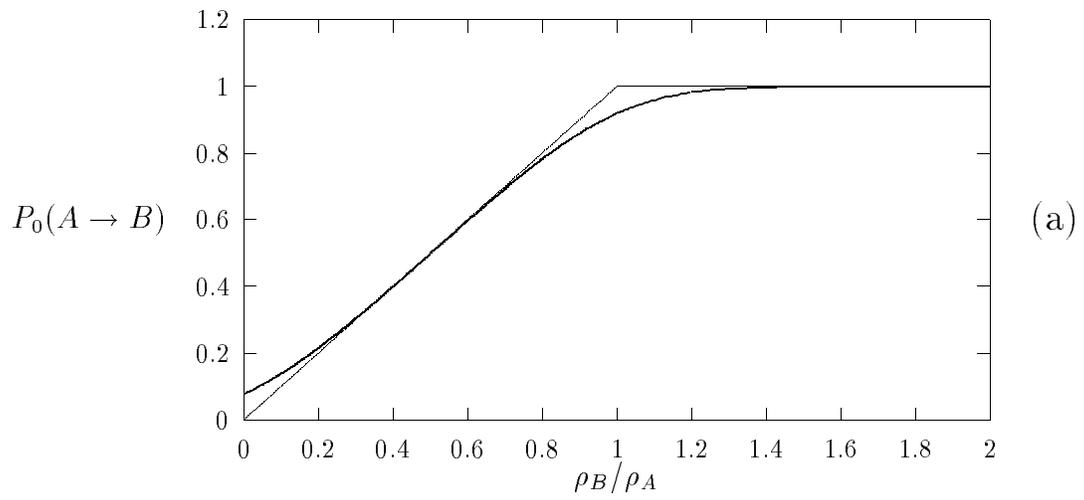
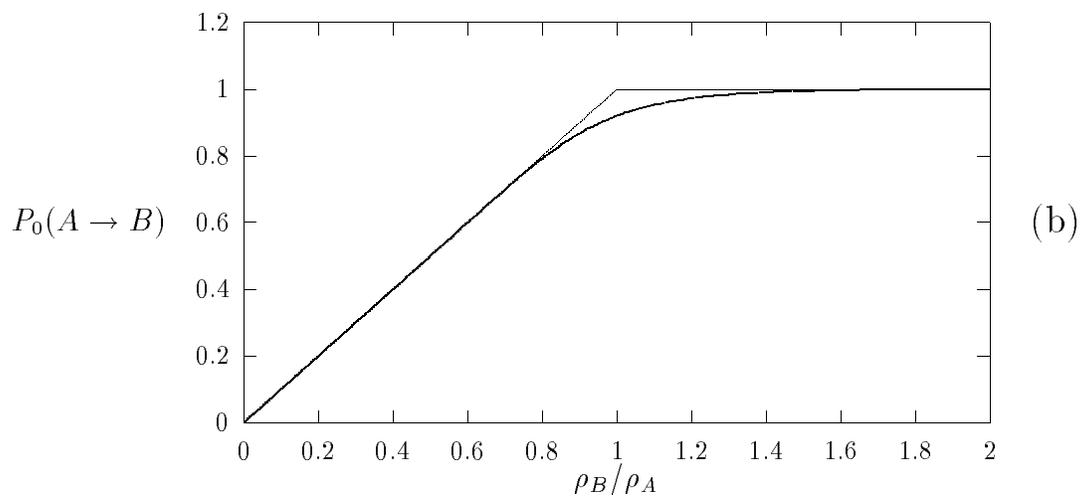
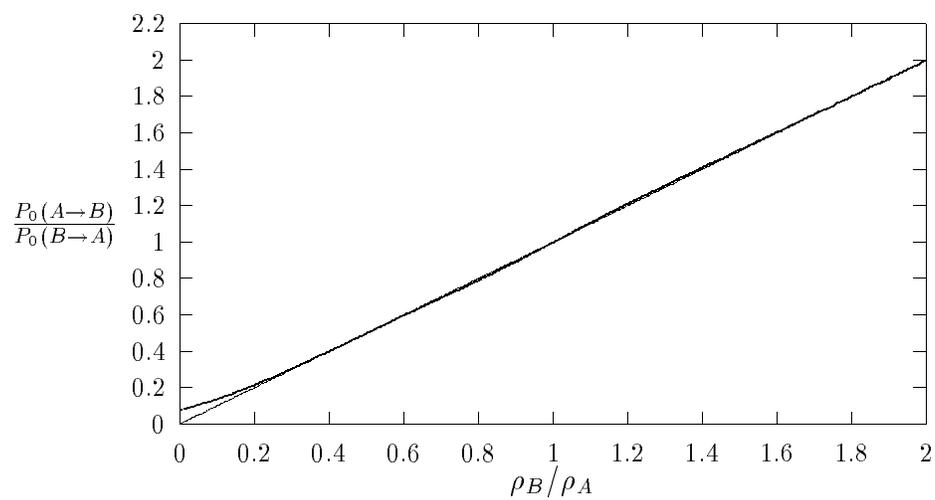
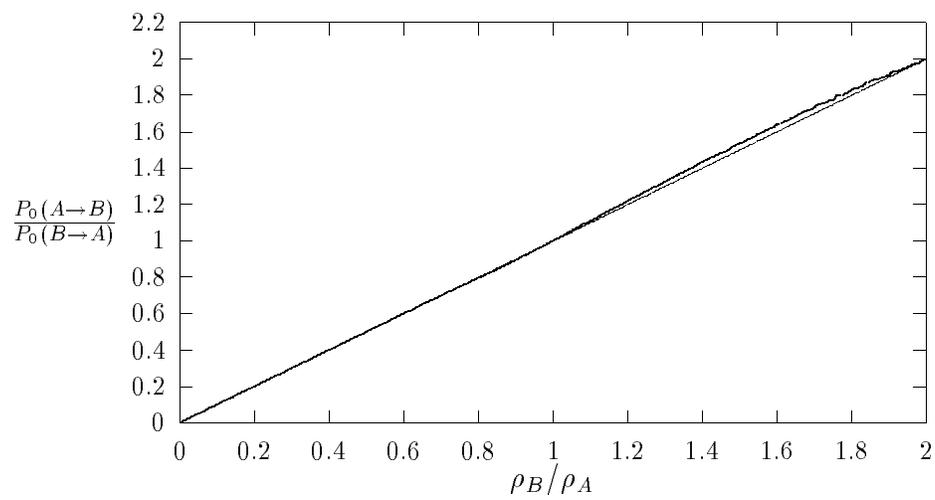
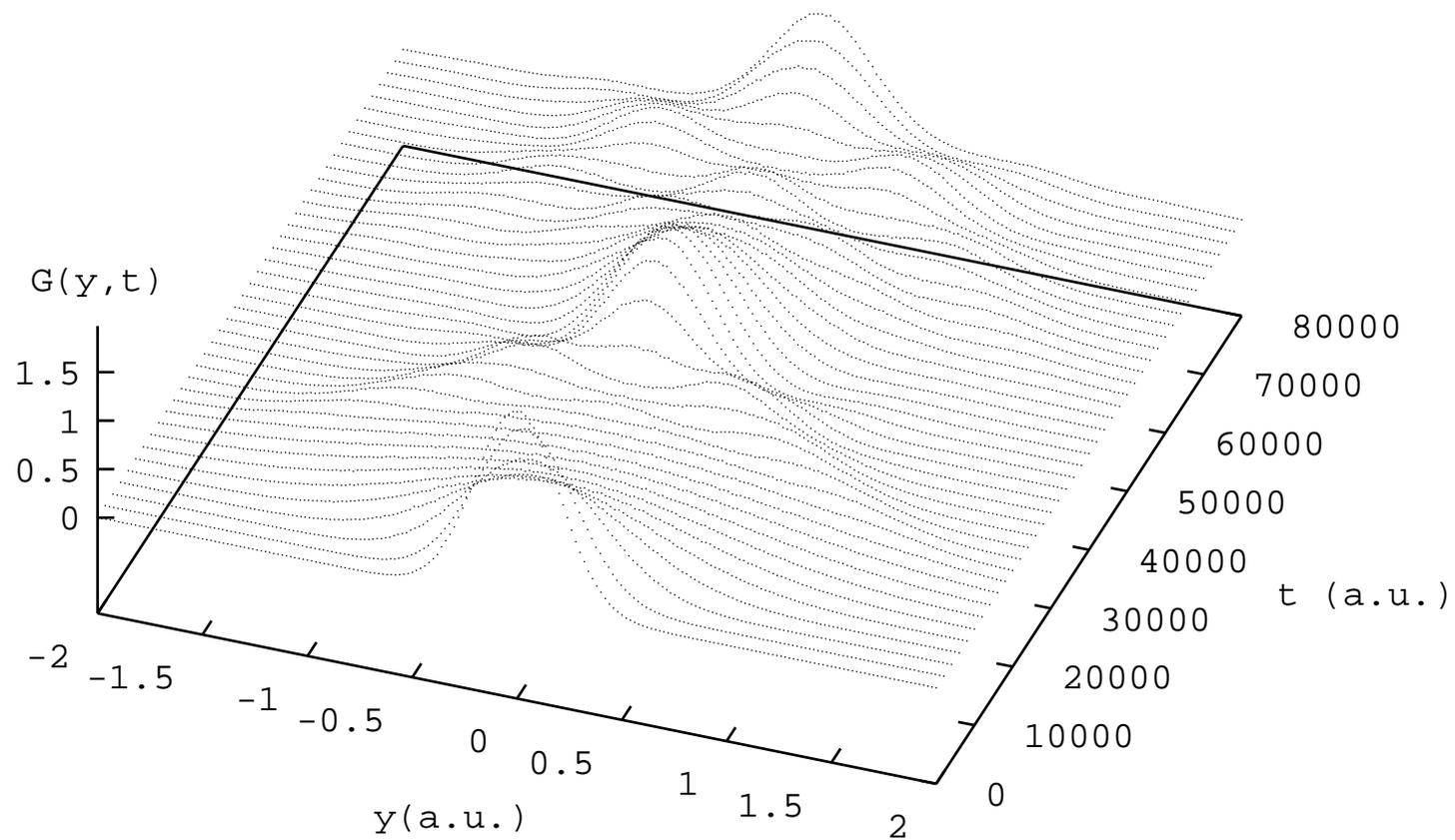
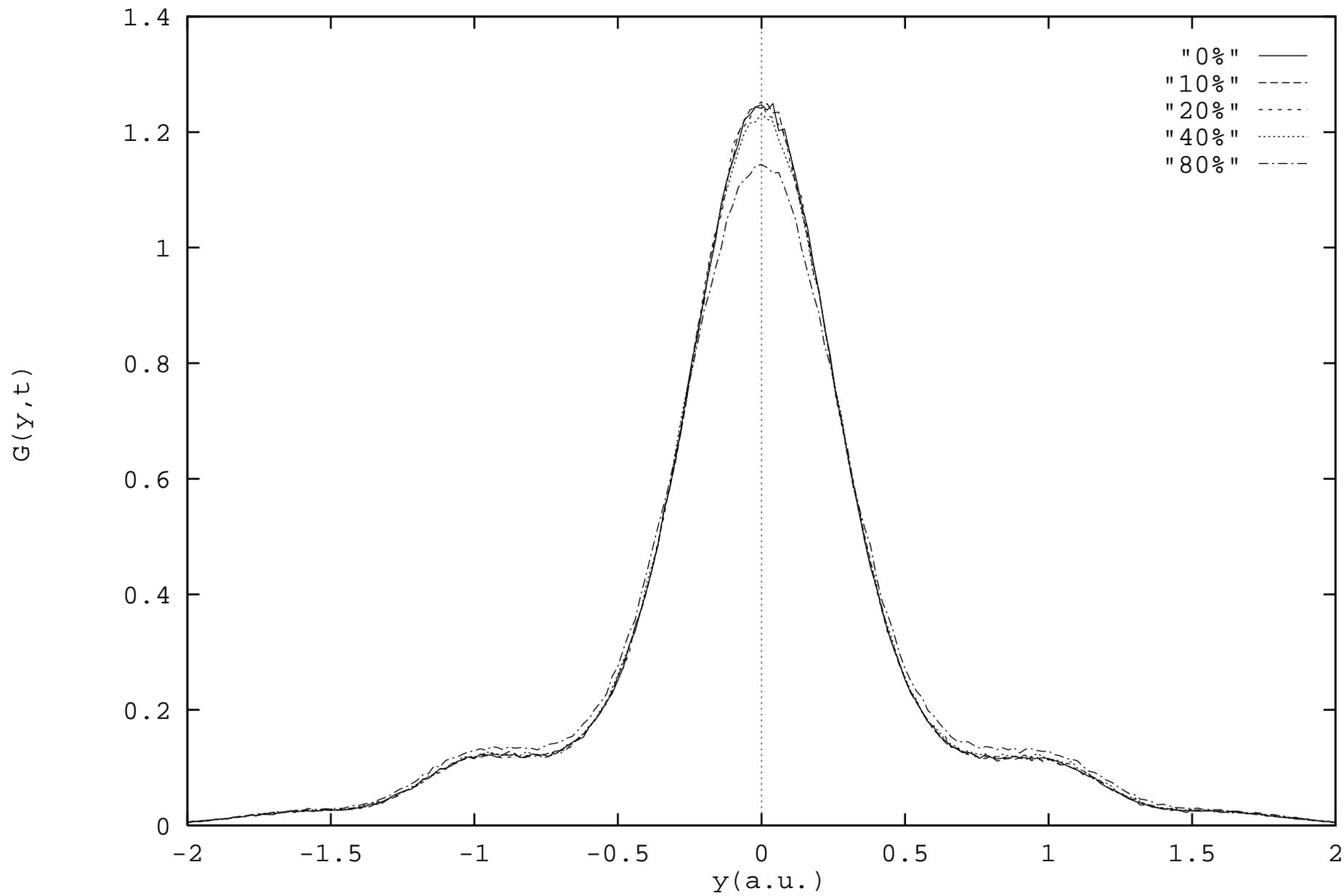
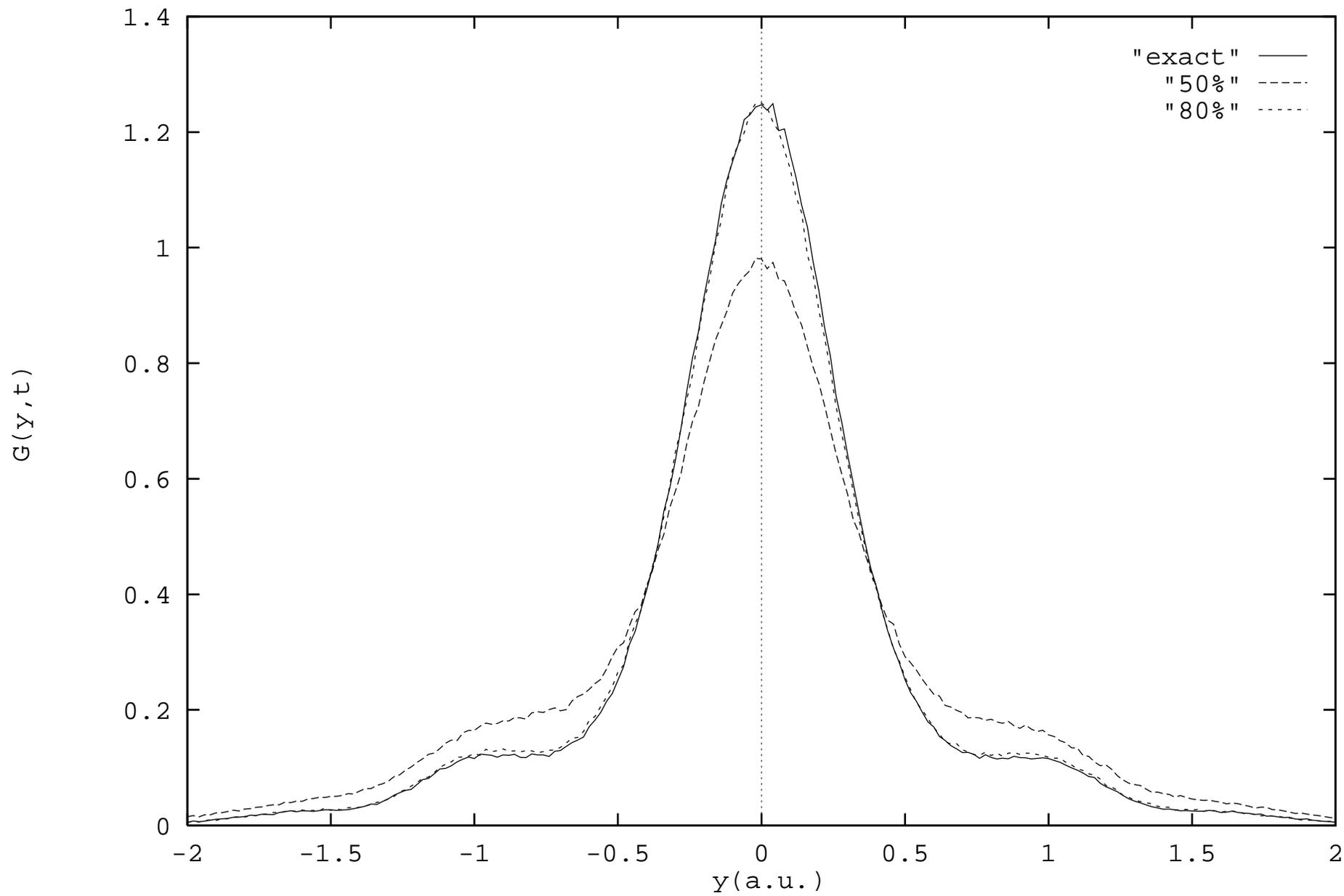
\end{document}